\documentclass{llncs} %

\usepackage{float}
\usepackage{graphicx}

\usepackage{wrapfig}
\usepackage{rotating}
\usepackage{epstopdf}
\usepackage{lscape}

\usepackage[labelfont=bf]{caption}
\usepackage[colorlinks=true,linkcolor=blue,citecolor=blue]{hyperref}

\usepackage{color,soul}
\usepackage{amsmath} 
\usepackage{mathtools} 
\usepackage{amsfonts} 
\usepackage{amssymb}
 \usepackage{rotating}
\usepackage{array} 
\usepackage{booktabs}
\usepackage{rotating}
\usepackage{multirow}
\usepackage{multicol}
\usepackage{lscape}
\usepackage{subfig}

\usepackage[export]{adjustbox}

\usepackage[ruled,vlined]{algorithm2e}

\usepackage{xargs}                      

\usepackage[colorinlistoftodos,prependcaption,textsize=tiny]{todonotes}

\usepackage{tikz,xcolor,hyperref}

\definecolor{lime}{HTML}{A6CE39}
\DeclareRobustCommand{\orcidicon}{
	\begin{tikzpicture}
	\draw[lime, fill=lime] (0,0) 
	circle [radius=0.16] 
	node[white] {{\fontfamily{qag}\selectfont \tiny ID}};
	\draw[white, fill=white] (-0.0625,0.095) 
	circle [radius=0.007];
	\end{tikzpicture}
	\hspace{-2mm}
}

\foreach \x in {A, ..., Z}{\expandafter\xdef\csname orcid\x\endcsname{\noexpand\href{https://orcid.org/\csname orcidauthor\x\endcsname}
			{\noexpand\orcidicon}}
}
			
\definecolor{darkgreen}{rgb}{0.53, 0.66, 0.42}

\begin{document}

\title{Predicting Brain Multigraph Population From a Single Graph Template for Boosting One-Shot Classification}

\titlerunning{Predicting Brain Multigraph Population From a Single Graph Template}  

\author{Furkan Pala\index{Pala, Furkan}\inst{1}, \and Islem Rekik\orcidA{} \index{Rekik, Islem}\inst{1}\thanks{ {corresponding author: irekik@itu.edu.tr, \url{http://basira-lab.com}.  }}}

\institute{BASIRA Lab, Faculty of Computer and Informatics Engineering, Istanbul Technical University, Istanbul, Turkey (\url{http://basira-lab.com/})}

\authorrunning{F. Pala et al.}

\maketitle              

\begin{abstract}
A central challenge in training one-shot learning models is the limited representativeness of the available shots of the data space. Particularly in the field of network neuroscience where the brain is represented as a graph, such models may lead to low performance when classifying brain states (e.g., typical vs. autistic). To cope with this, most of the existing works involve a data augmentation step to increase the size of the training set, its diversity and representativeness. Though effective, such augmentation methods  are limited to generating samples with the same size as the input shots (e.g., generating brain connectivity matrices from a single shot matrix). To the best of our knowledge, the problem of generating brain multigraphs capturing multiple types of connectivity between pairs of nodes (i.e., anatomical regions) from a single brain graph remains unsolved. In this paper, we unprecedentedly propose a hybrid graph neural network (GNN) architecture, namely Multigraph Generator Network or briefly MultigraphGNet, comprising two subnetworks: (1) \emph{a many-to-one GNN} which integrates an input population of brain multigraphs into a single template graph, namely a connectional brain temple (CBT), and (2) \emph{a reverse one-to-many} U-Net network which takes the learned CBT in each training step and outputs the reconstructed input multigraph population. Both networks are trained in an end-to-end way using a cyclic loss. Experimental results demonstrate that our MultigraphGNet boosts the performance of an independent classifier when trained on the augmented brain multigraphs in comparison with training on a single CBT from each class. We hope that our framework can shed some light on the future research of multigraph augmentation from a single graph. Our MultigraphGNet source code is available at \url{https://github.com/basiralab/MultigraphGNet}.

\end{abstract}


\section{Introduction}
Brain graphs present powerful tools in modeling the relationship between different anatomical regions of interest (ROIs) as well as fingerprinting neural states (e.g., typical and atypical) \cite{van2019}. Recently, graph neural network (GNN) models have achieved remarkable results across different brain graph learning tasks \cite{bessadok2021} such as time-dependent prediction \cite{tekin2021recurrent,gurler2020foreseeing}, super-resolution \cite{isallari2021,mhiri2021IPMI} and classification \cite{oh2022diagnosis,nebli2022quantifying}. Despite their ability to extract meaningful and powerful representations from labelled brain graph data, they might fail to handle training data with a limited number of samples. Particularly, such data-hungry architectures might struggle to converge and produce a good performance within a few-shot learning (FSL) paradigm \cite{kadam2018review,sun2019meta,li2020concise} --let alone one-shot learning \cite{guvercin2021one}. 

Such problem is usually remedied by data augmentation where labeled samples are generated from the available shots to better generalize to unseen distributions of testing samples. Several FSL works \cite{kotia2021few} proposed novel methods to solve medical image-based learning tasks.  For instance, \cite{zhao2019data} presented a learning-based method that is trained on a few samples while leveraging data augmentation and unlabeled image data to enhance model generalizability. \cite{li2020difficulty} used the meta-train data from common diseases for rare disease diagnosis and tackled the low-data regime problem while leveraging meta-learning. \cite{chaitanya2021} presented a novel task-driven and semi-supervised data augmentation scheme to improve medical image segmentation performance in a limited data setting. However, to the best of our knowledge and as revealed by this recent GNN in network neuroscience review paper \cite{bessadok2021}, one-shot GNN learning remains unexplored in the field of network neuroscience --with the exception of \cite{guvercin2021one} where one-shot GNN architectures are trained for brain connectivity regression and classification tasks. Specifically, representative  connectional brain templates (CBTs) \cite{chaari2022comparative} were used to train GNN architectures in one-shot fashion. Such graph templates present a compact representation of a particular brain state.
 
However, this landmark work did not resort to any data augmentation strategies or generative models to better estimate the unseen distributions of the classes to discriminate. Besides, existing graph augmentation methods are limited to generating graphs with the same size as the input shots (e.g., generating brain connectivity matrices from a single shot matrix). To the best of our knowledge, the problem of generating brain multigraphs capturing multiple types of connectivity between pairs of nodes (i.e., anatomical regions) from a single brain graph remains unsolved. Note that a brain multigraph is encoded in a tensor, where each frontal view captures a particular type of connectivity between pairs of brain ROIs (e.g., morphological or functional). In this paper, we set out to boost a one-shot brain graph classifier \emph{by learning how to generate multi-connectivity brain multigraphs from a single template graph}. Specifically, we propose a hybrid graph neural network (GNN) architecture, namely Multigraph Generator Network or briefly MultigraphGNet, comprising two subnetworks: (1) \emph{a many-to-one GNN} which integrates an input population of brain multigraphs into a single CBT graph using deep graph normalizer (DGN) \cite{gurbuz2020}, and (2) \emph{a reverse one-to-many convolutional neural network (CNN)} which takes the learned CBT in each training step and outputs the reconstructed input multigraph population. Our prime contributions are listed below:
\begin{enumerate}
    \item We are the first to learn how to generate brain multigraphs from a single graph template (namely CBT).
    \item We propose a  hybrid cyclic GNN architecture for multigraph graph augmentation from a single CBT.
    \item We show that the augmented brain multigraphs can boost the performance of an independent classifier across various evaluation metrics. 
\end{enumerate}

\section{Methodology}

In this section, we explain our proposed MultigraphGNet in detail. We represent tensors by calligraphic font capital letters, e.g., $\mathcal{X}$, matrices by boldface capital letters, e.g., $\mathbf{X}$, vectors by boldface lowercase letters, e.g., $\mathbf{x}$ and scalars by letters, e.g., $x$. Table~\ref{tab:1} summarizes the mathematical notations we used throughout the paper.
\begin{table}
\caption{Mathematical notations followed in the paper}\label{tab:1}
\begin{tabular}{c|c}
\hline
Mathematical notation &  Definition \\
\hline
$S$ & Training set \\
$n_r$ & Number of region of interests (ROIs) in the brain \\
$n_v$ & Number of connectomic views in the brain multigraph (tensor) \\
$\mathcal{X}_s$ & Brain graph tensor $\in \mathbb{R}^{n_r \times n_r \times n_v}$ of subject $s$\\
$\mathbf{X}_s^v$ & Brain graph matrix $\in \mathbb{R}^{n_r \times n_r}$ of the view $v$ and subject $s$\\
$\mathbf{C}_s$ & Subject-driven connectional brain template (CBT) $\in \mathbb{R}^{n_r \times n_r}$ of the subject $s$\\
$\hat{\mathcal{X}}_s$ & Reconstructed brain graph tensor $\in \mathbb{R}^{n_r \times n_r \times n_v}$ for the subject $s$\\
\hline
\end{tabular}
\end{table}

\begin{sidewaysfigure}
\centering
\includegraphics[width=19cm]{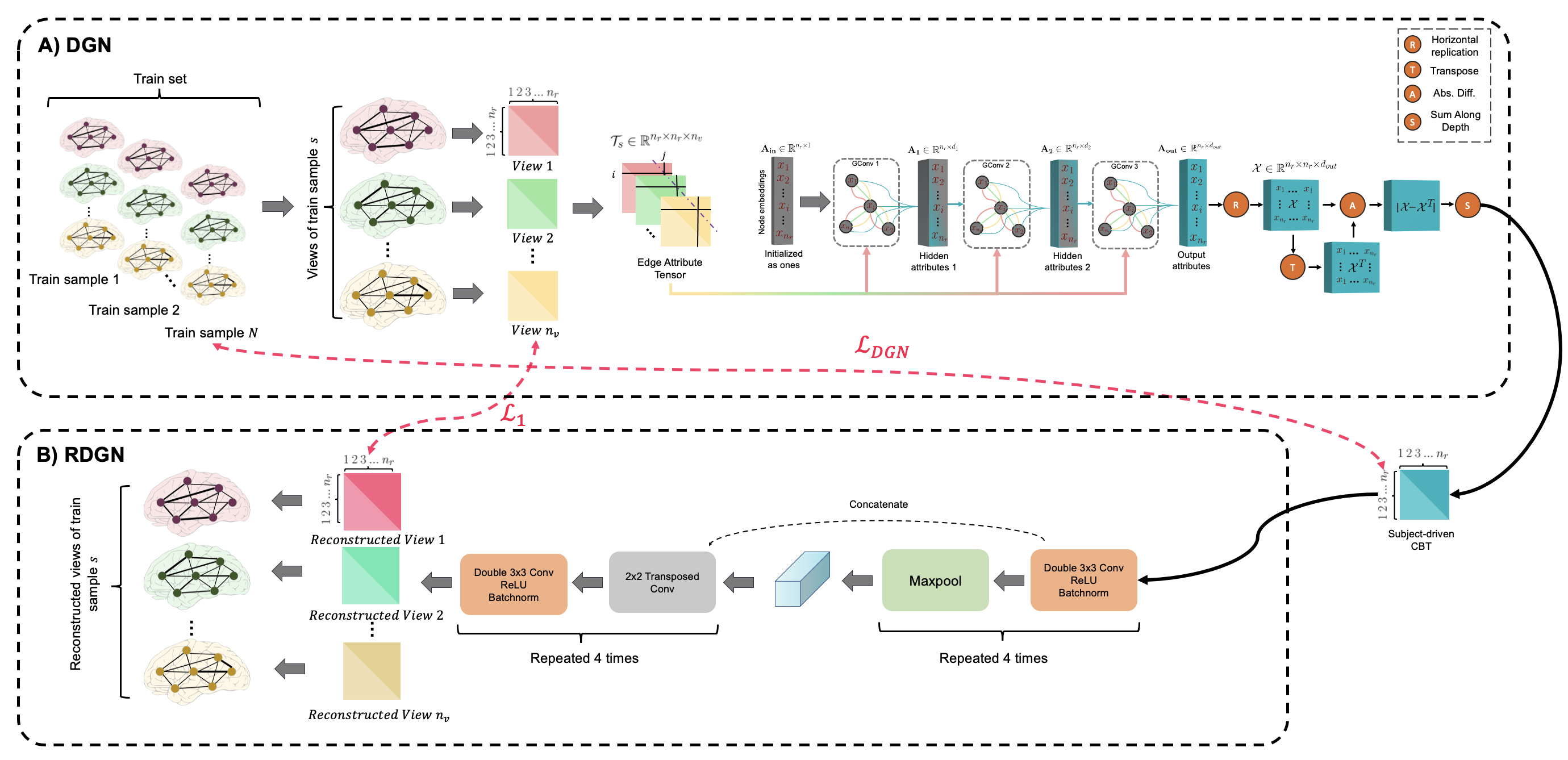}
\caption{\emph{The proposed MultigraphGNet architecture to predict a population of brain multigraphs from a single graph template shot.} \textbf{A) CBT generation from a population of brain multigraphs.} To represent the multiple views of a brain graph in a single view, Deep Graph Normalizer (DGN) \cite{gurbuz2020} network learns the node embeddings using three edge conditioned graph convolutional layers followed by differentiable tensor operations to generate a subject specific CBT. DGN Loss computes the mean Frobenius distance between the CBT and each view in the training subjects. \textbf{B) Reverse brain multigraph reconstruction from the learned CBT.} To reconstruct the original views of the training subject, we propose a novel Reverse-DGN (RDGN) network that uses a U-Net architecture to first encode the CBT into a lower-dimensional representation using the same convolution and maxpool layers in DGN, then up-sample to multiple views via transposed convolution layers. We optimize the reconstructed tensor population by minimizing the L1 distance between the original and reconstructed views for each training subject. DGN and RDGN are trained in a cyclic and end-to-end manner to ensure that (i) the learned CBT captures the connectivity patterns shared across brain views and subjects and (ii) the tensor views of each training subject can be reconstructed using the learned CBT solely.} 
\label{fig:main_fig}
\end{sidewaysfigure}

\textbf{Problem statement.} A brain connectome can be encoded in a single view (i.e., matrix) or multiple views (i.e., matrices forming a tensor) so that each view sits on a different manifold and captures a specific relationship, e.g., morphological or functional, between anatomical brain regions of interest. Since multiview connectomic data is scarce, we set out to learn how to predict brain connectivity tensors (i.e., multigraphs) from a single graph template (i.e., brain connectivity matrix). Thus, we propose a one-to-many brain graph augmentation approach. Specifically, given a set of multi-view brain graphs where each view models a specific relationship between pairs of brain ROIs, our goal is to first collapse these graphs to a single view graph-based representation, i.e., connectional brain template (CBT), then, reconstruct the original brain graphs using the generated CBT, so that we can augment new multi-view brain graphs by adding small noise to the global CBT which can be considered as an average connectome over all subjects and views. 

\textbf{Definition 1.} Let $\mathbf{C}_s$ denote a subject-driven connectional brain template, which is a centered representation of subject $s$ with respect to the training population tensor (i.e., multigraph) distribution. Specifically, $\mathbf{C}_s$ is encoded in a single-view brain connectivity matrix which is a normalized graph-based representation of the multi-view brain graph (i.e., tensor) of subject $s$. 

\subsection{CBT Learning}
The first block (\textbf{Fig.}~\ref{fig:main_fig}-A) of our MultigraphGNet utilizes the Deep Graph Normalizer (DGN)~\cite{gurbuz2020} network to produce a subject-driven CBT $\mathbf{C}_s$ for each training subject $s$. We represent a brain graph view $i$ as ${G_i({V_i},{E_i})}$ where $V_i$ is the set of $n_r$ nodes each corresponds to a brain ROI and $E$ is a set of edges each encoding a particular type of relationship between two ROIs (e.g., structural). Thus, we can define a multi-view brain graph for subject $s$ as a tensor $\mathcal{X}_s \in \mathbb{R}^{n_r \times n_r \times n_v}$, where $n_r$ and $n_v$ denote the number of ROIs and views, respectively. Since self-connections do not carry important information, we set the diagonal entries in the tensor to zero. The DGN network takes a node embedding matrix $\mathbf{V^0} \in \mathbb{R}^{n_r \times d_0}$, where $d_0$ is the initial node embedding size and a multi-view edge embedding tensor $\mathcal{X}$. Since we do not have any node/ROI features initially, we set the $\mathbf{V^0}$ to $\mathbf{1}$. DGN utilizes 3 edge-conditioned graph convolution layers~\cite{simonovsky2017} with a ReLU at the end of each layer to learn the node embeddings. Each layer $l \in \{1,2,3\}$ includes a dense filter neural network $F^l: \mathbb{R}^{n_v} \mapsto \mathbb{R}^{d_l \times d_{l-1}}$ that implements the message passing between ROIs $i$ and $j$ given the edge embeddings $\mathbf{e}_{ij} \in \mathbb{R}^{n_v \times 1}$ as follows
\[
    \mathbf{v}_i^l = \mathbf{\Theta}^l \mathbf{.} \mathbf{v}_i^{l-1} + \frac{1}{|\mathcal{N}(i)|} \Biggl( \sum_{j \in \mathcal{N}(i)} F^l(\mathbf{e}_{ij};\mathbf{W}^l)\mathbf{v}_j^{l-1} + \mathbf{b}^l \Biggl)
\]
\[
    F^l(\mathbf{e}_{ij};\mathbf{W}^l) = \mathbf{\Theta}_{ij},
\]
where $\mathbf{v}_i^l$ is the node embedding corresponding to the $i^{th}$ ROI in layer $l$. The dense filter neural network $F^l$ with weights $\mathbf{W}^l$ and bias $\mathbf{b}^l$ produces new edge weights in each layer for the edges between node $i$ and its neighbour $j \in \mathcal{N}(i)$. The resulting node embeddings tensor $V^3 \in \mathbb{R}^{n_r \times d_3}$ is first repeated horizontally to get a tensor, then, we compute the element-wise absolute difference between its transpose. Here we use the absolute difference since the original brain connectivity tensors were generated using this operation. One can use any other operation that is differentiable for the back propagation process. The final output is obtained by summing along the $z$-axis which gives us the subject-driven CBT $\mathbf{C}_s \in \mathbb{R}^{n_r \times n_r}$ for subject $s$.

We use the Subject Normalization Loss (SNL) as proposed in the DGN. SNL for training subject $s$ is defined as the mean Frobenius distance between the learned CBT and each training subject view as follows:
\[
    SNL_s = \frac{1}{n_v \times |S|}\sum_{v=1}^{n_v}\sum_{i \in S} ||\mathbf{C}_s - \mathbf{X}_i^v||_F \times \lambda_{v},
\]
where $S$ is the training set and $\lambda_v$ is the loss weight computed for each view $v$ as follows:
\[
    \lambda_v = \frac{\frac{1}{\mu_v}}{\text{max}\Bigl\{ \frac{1}{\mu_j} \Bigl\}_{j=1}^{n_v}},
\]
where $\mu_v$ is computed by taking the mean of edge attributes for view $v$. Next, we define and optimize the DGN Loss using the following objective function \cite{gurbuz2020}:
\[
    \mathcal{L}_{\text{DGN}} = \frac{1}{|S|}\sum_{s \in S}SNL_s
\]

\subsection{Reverse Mapping}
The second block (\textbf{Fig.}~\ref{fig:main_fig}-B) of our MultigraphGNet aims to reverse the DGN process, thus we call it reverse DGN (RDGN) network by taking the learned CBT $\mathbf{C}_s$ and mapping it back into the original brain multigraph tensor $\mathcal{X}_s$ for subject $s$.  We use the U-Net~\cite{ronneberger2015} architecture to design the RDGN, which consists of an encoder and a decoder. Specifically, in each iteration of the optimization process, the encoder takes the learned CBT and applies the same convolution operation with a kernel size of 3, stride and padding of 1 two times, each followed by a ReLU non-linearity and a batch normalization layer. To down-sample the resulting feature map, we use a max pooling layer with a kernel size and stride of 2. The number of output channels is doubled at the end of the down-sampling. We repeat this process 4 times to get the feature map with 1024 channels. In the decoder part, we first up-sample using a 2$\times$2 transposed convolutional layer with a stride of 2 that halves the number of output channels which is followed by a concatenation with the feature map from the counterpart in the encoder. As for the decoder, we apply the same convolution operation twice. This process is repeat 4 times, as well. The final layer consists of a 1$\times$1 convolution layer that outputs the reconstructed tensor $\hat{\mathcal{X}}_s \in \mathbb{R}^{n_r \times n_r \times n_v}$. To preserve the similarity to the original tensor, we additionally minimize the $L1$ distance, i.e., mean absolute error (MAE), between the original ($\mathcal{X}$) and reconstructed ($\hat{\mathcal{X}}$) tensor views as follows:
\[
    \mathcal{L}_{L1} = ||\mathcal{X} - \hat{\mathcal{X}}||_1 
\]
We train the DGN and RDGN in an end-to-end and fully cyclic manner to ensure that the generated CBT can well collapse the multiple views into a single connectivity matrix, which in turn is used to reconstruct back the original brain multigraph using the U-Net augmentation process. Thus, we define our RDGN loss as follows:
\[
    \mathcal{L}_{cyclic} = \mathcal{L}_{DGN} + \lambda \mathcal{L}_{L1}
\]

\subsection{Multigraph data augmentation from a single graph}\label{sec:aug}
The trained RDGN is able to generate multiple views from a given single-view CBT, which makes it possible to predict a multigraph from a single representative template graph. Hence, by slightly modifying the an particular input CBT, RDGN can produce unique brain multigraphs where it acts as a \emph{one-to-many augmentation network}. To augment new brain multigraphs from a single CBT, we first obtain a subject-driven CBT for each training subject as follows:
\[
    S_{CBT} = \Bigl\{ \mathbf{C}_s | \forall s \in S \Bigl\}
\]
\[
    \mathbf{C}_s = DGN(\mathcal{X}_s)
\]
Next, we construct a global CBT $\mathbf{C}$ by taking the element-wise median of all the subject-driven CBTs in $S_{CBT}$. $\mathbf{C}$ can be considered as an average brain network over all the training brain multigraph set. To create diversity in our augmented multigraphs, we add a small noisy matrix $\mathbf{W}_i \sim \mathcal{N}(\mu_{\mathbf{C}}, \sigma_{\mathbf{C}}^2)$ to the global CBT for each augmented new sample $i$ as follows:
\[
    \Tilde{\mathbf{C}}_i = \mathbf{C} + c\mathbf{W}_i,
\]
where $\mu_{\mathbf{C}}$ denotes the CBT mean and $\sigma_{\mathbf{C}}$ its standard deviation. $c$ is a scaling coefficient to control the added noise. Note that we re-sample the added noise in each augmentation step. We augment new samples as follows:

\[
    S_{aug} = \Bigl\{ \hat{\mathcal{X}}_i | i \in \{ 1,2,\dots,k \} \Bigl\}
\]
\[
    \hat{\mathcal{X}}_i = RDGN(\Tilde{\mathbf{C}}_i),
\]
where $k$ is the number of samples that we want to augment.

\section{Experimental results and discussion}

\textbf{Evaluation dataset.} We evaluated our framework on the Autism Brain Imaging Data Exchange (ABIDE-I) public dataset\footnote{\url{http://preprocessed-connectomes-project.org/abide/}} using a random subset including 150 normal control (NC) and 150 subjects with autism spectrum disorder (ASD), each wit 6 views of morphological brain connectomes (extracted from the maximum principal curvature, the mean cortical thickness, the mean sulcal depth, the average curvature, the minimum principle area and the cortical surface area) of the left cortical hemispheres (LH). The cortical surface is split into 35 ROIs via Desikan-Killiany atlas~\cite{fischl2004} after the reconstruction from T1-weighted MRI using the FreeSurfer pipeline~\cite{fischl2012}. Next, the brain network is obtained by taking the absolute difference between the cortical measurements in each pair of ROIs. We used 5-fold cross-validation with 5 different seeds to evaluate the generalizability of our MultigraphGNet. We implemented our framework in PyTorch and PyTorch-Geometric~\cite{fey2019} libraries.

\textbf{Hyperparameters.} In DGN, we used 3 edge-conditioned graph convolution layers followed by ReLU non-linearity and each layer has an output node embedding size of 36, 24 and 5, respectively. In RDGN, we used a U-Net architecture. For the optimizer, we chose AdamW~\cite{loshchilov2018} with a learning rate of 0.001, beta1 and beta2 of 0.9 and 0.999, and a weight decay of 0.01.  We set $\lambda =1$ in the total loss function and $c = 0.2$ for the added noise in the CBT augmentation using RDGN.

\begingroup
\renewcommand{\arraystretch}{1.3}
\begin{table}[htp!]
\centering
\caption{Testing classification results of independent SVM classifiers trained using (i) a single CBT from each class and (ii) samples augmented using the trained RDGN network. We report the average accuracy, precision, recall and F1 score obtained when training on 10, 25 and 50 augmented samples. Each row displays the mean of the results over the 5 cross-validation folds with different random seeds for the train-test split.}
\label{tab:classification_results}
\begin{tabular}{c|cccc|cccc}
\hline
\multirow{2}{*}{} & \multicolumn{4}{c|}{One-shot CBT}      & \multicolumn{4}{c}{Augmented multigraphs} \\ \cline{2-9} 
                  & Acc   & Prec           & Rec   & F1    & Acc              & Prec   & Rec             & F1              \\ \hline
Seed \#1          & 0.929 & 0.944          & 0.916 & 0.928 & 0.989            & 0.953  & 0.993           & 0.989           \\
Seed \#2          & 0.960 & 0.993          & 0.928 & 0.959 & 0.956            & 0.939  & 1.000           & 0.964           \\
Seed \#3          & 0.948 & 0.993          & 0.903 & 0.944 & 0.971            & 0.985  & 0.955           & 0.968           \\
Seed \#4          & 0.961 & 0.981          & 0.941 & 0.960 & 0.965            & 0.991  & 0.935           & 0.958           \\
Seed \#5          & 0.954 & 0.992          & 0.915 & 0.952 & 0.968            & 0.961  & 0.979           & 0.969           \\ \hline
Avg.              & 0.950 & \textbf{0.981} & 0.921 & 0.949 & \textbf{0.970}   & 0.966  & \textbf{0.972}  & \textbf{0.970}  \\ \hline
\end{tabular}
\end{table}
\endgroup

\begin{figure}[ht!]
\centering
\includegraphics[width=12cm]{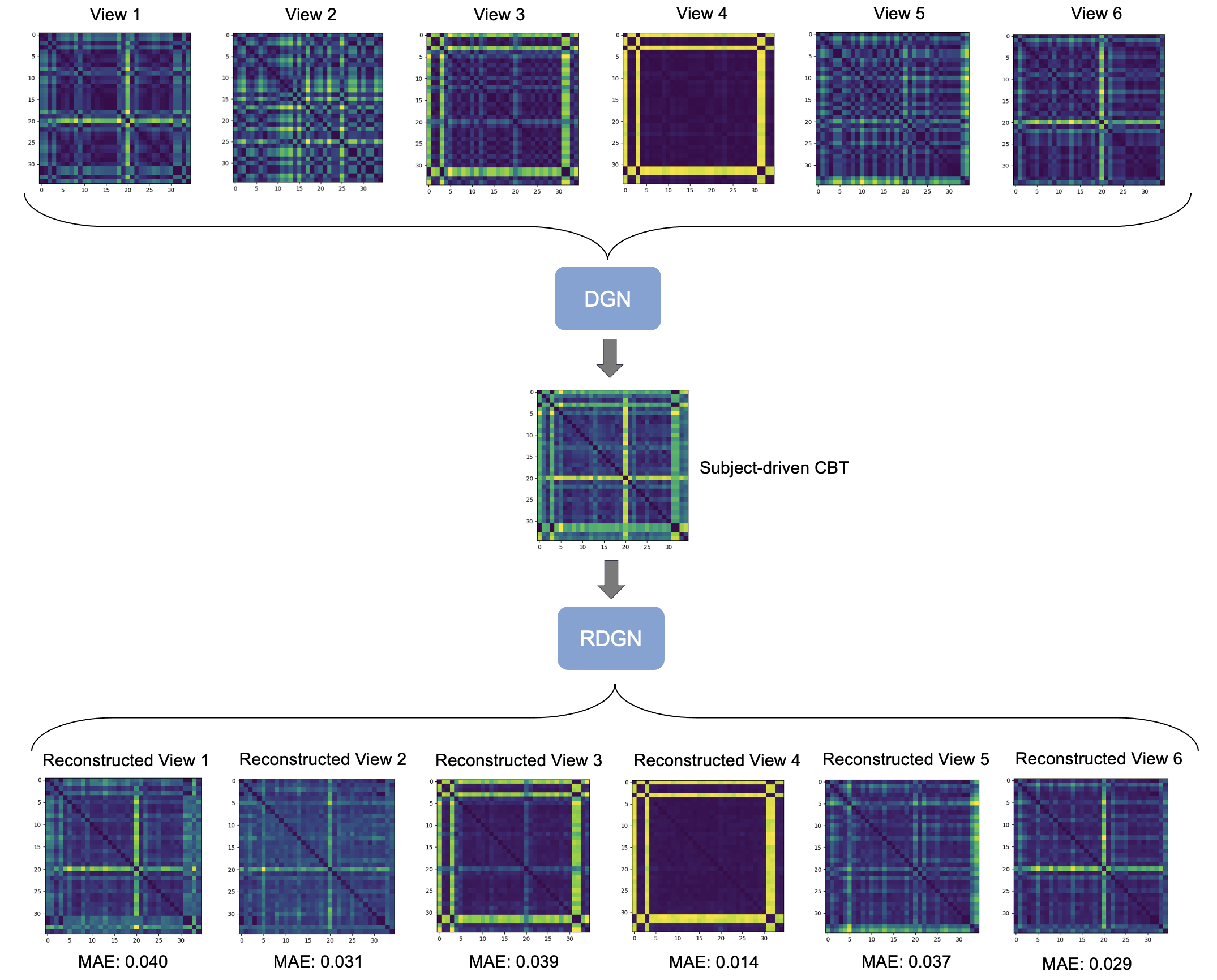}
\caption{Visual inspection of the reconstructed brain multigraph from a single CBT. On top, we display the brain tensor including 6 connectivity views for a randomly selected ASD testing sample. In the middle, we present the subject-driven CBT obtained using the trained DGN network. In the bottom, we compare the reconstructed views by the trained RDGN network given the learned CBT as input. We also measure the MAE between the corresponding ground-truth and predicted views.} 
\label{fig:visual_inspection}
\end{figure}

\textbf{Evaluation and comparison methods.} To evaluate the effectiveness of our brain multigraph augmentation strategy from a single CBT, we trained two support vector machine (SVM) classifiers in each cross-validation fold. Note that we provided the same seeds in both DGN/RDGN and SVM training so that neither our augmentation framework nor the classifier have seen the test set before. 

\emph{One-shot CBT.} We generated two global CBTs $\mathbf{C}_{ASD}$ and $\mathbf{C}_{NC}$ using the trained DGN for ASD and NC training sets, respectively. We vectorized the upper-triangular part of both CBTs to get two feature vectors $\mathbf{c}_{ASD},\mathbf{c}_{NC} \in \mathbb{R}^{\frac{n_r \times n_r - 1}{2}}$. In the testing step, we created a subject-driven CBT using the trained DGN for each brain multigraph in the test set since the SVM was not trained on multiple views.

\emph{Augmented samples.} We augmented $k=10,25,50$ multigraphs as explained in the Section~\ref{sec:aug} and vectorized the upper-triangular parts of each view to get the feature vector $\mathbf{x}_i \in \mathbb{R}^{\frac{n_v \times n_r \times n_r - 1}{2}}$ for each augmented sample $i \in \{1,2,\dots,k \}$. Next, we trained a new SVM classifier on the augmented set and tested it on the left-out test set. In this case, there is no need for generating subject-driven CBTs in the testing phase since the SVM was already trained on multi-view tensors. We report the comparison between the classification accuracy, precision, recall and F1 scores for the both methods in the Table~\ref{tab:classification_results} for both methods. It can be clearly seen that our framework is able to reconstruct the initial brain graph views and produces relevant features for the ASD/NC classification task. While one-shot CBT is considerably enough to distinguish between ASD and NC subjects, our framework further boosts the independent classifier performance by augmenting multiple multi-view brain connectomes using only one single-view CBT.

\textbf{Visual inspection.} In \textbf{Fig.}~\ref{fig:visual_inspection}, we show the original and reconstructed brain multigraph tensor including 6 views as well as the learned CBT for a randomly selected ASD testing subject. In addition, we report the mean absolute error (MAE) between reconstructed and original views. Obviously, RDGN network is able to expand and decode the CBT into multiple views with a low error.

\textbf{Limitations and future directions.} Although the L1 Loss between the the ground truth and reconstructed views produced very promising reconstructions and is resistant to data outliers, it only considers the element-wise similarity in connectivity weights without examining the topological properties (e.g., hubness) of the augmented multigraphs. Hence in our future work, we will add a topological sub-loss to the cyclic loss in our reconstruction block. Furthermore, the RDGN can be further boosted by adding a discriminator network to the U-Net under an adversarial learning paradigm as in \cite{gurler2020foreseeing}. We also will replace the convolutional U-Net with a graph U-Net \cite{gao2019graph} for further improvement.

\section{Conclusion}

In this paper, we introduced  the first study that provides a one-to-many U-Net augmentation framework for generating multi-view brain graphs from a single connectional template to boost one-shot learning classifiers. Given the high-cost of connectomic data collection and processing, our framework offers an affordable approach to learning how in a frugal setting with limited data. We showed that the augmented samples are able to improve the classification results of autistic subjects. In our future work, we will evaluate our MultigraphGNet on subjects with different neurological disorders such as Alzheimer's Disease (AD) or mild cognitive impairment (MCI) and assess the generalizability of  model to different classes.

\section{Supplementary material}

We provide three supplementary items for reproducible and open science:

\begin{enumerate}
	\item A 7-mn YouTube video explaining how our framework works on BASIRA YouTube channel at \url{https://youtu.be/LQZBVwo_iNU}.
	\item MultigraphGNet code in Python on GitHub at \url{https://github.com/basiralab/MultigraphGNet}. 
	\item A GitHub video code demo on BASIRA YouTube channel at \url{https://youtu.be/iNNFNlML_CU}. 
\end{enumerate}

\section{Acknowledgements}

This work was funded by generous grants from the European H2020 Marie Sklodowska-Curie action (grant no. 101003403, \url{http://basira-lab.com/normnets/}) to I.R. and the Scientific and Technological Research Council of Turkey to I.R. under the TUBITAK 2232 Fellowship for Outstanding Researchers (no. 118C288, \url{http://basira-lab.com/reprime/}). However, all scientific contributions made in this project are owned and approved solely by the authors.

\bibliography{Biblio3}
\bibliographystyle{splncs}
\end{document}